\pdfoutput=1

\documentclass[11pt]{article}

\usepackage[final]{acl}

\usepackage{times}
\usepackage{latexsym}
\usepackage{graphicx}
\usepackage[T1]{fontenc}

\usepackage[utf8]{inputenc}
\usepackage{booktabs}

\usepackage{microtype}

\usepackage{inconsolata}

\usepackage{multirow}
\usepackage{multicol}
\usepackage{caption}
\usepackage{subcaption}
\usepackage{comment}

\title{AIR-Bench: Benchmarking Large Audio-Language Models via Generative Comprehension}

\author{%
  Qian Yang$^1$\thanks{~~Equal contribution.}\thanks{~~Intern at Alibaba.},  Jin Xu$^2$\footnotemark[1], Wenrui Liu$^1$, Yunfei Chu$^2$, Ziyue Jiang$^1$, Xiaohuan Zhou$^2$ \\ \textbf{Yichong Leng$^2$, Yuanjun Lv$^2$, Zhou Zhao$^1$\thanks{~~Corresponding to Zhou Zhao~(zhaozhou@zju.edu.cn) and Chang Zhou~(ericzhou.zc@alibaba-inc.com).}, Chang Zhou$^2$\footnotemark[3], Jingren Zhou$^2$}
  \\
  $^1$Zhejiang University, $^2$Alibaba Group \\
  \texttt{\{qyang1021,liuwenrui,ziyuejiang,zhaozhou\}@zju.edu.cn} \\
  \texttt{\{renjun.xj,fay.cyf,shiyi.zxh,lengyichong.lyc,lvyuanjun.lyj\}@alibaba-inc.com} \\
  \texttt{\{ericzhou.zc,jingren.zhou\}@alibaba-inc.com} 
}

\begin{document}
\maketitle
\begin{abstract}
Recently, instruction-following audio-language models have received broad attention for human-audio interaction. However, the absence of benchmarks capable of evaluating audio-centric interaction capabilities has impeded advancements in this field. Previous models primarily focus on assessing different fundamental tasks, such as automatic speech recognition, and lack an assessment of the open-ended generative capabilities centered around audio. Thus, it is challenging to track the progression in the Large Audio-Language Models (LALMs) domain and to provide guidance for future improvement.
In this paper, we introduce AIR-Bench (\textbf{A}udio \textbf{I}nst\textbf{R}uction \textbf{Bench}mark), the first benchmark designed to evaluate the ability of LALMs to understand various types of audio signals (including human speech, natural sounds, and music), and furthermore, to interact with humans in the textual format. AIR-Bench encompasses two dimensions: \textit{foundation} and \textit{chat} benchmarks. The former consists of 19 tasks with approximately 19k single-choice questions, intending to inspect the basic single-task ability of LALMs. The latter one contains 2k instances of open-ended question-and-answer data, directly assessing the comprehension of the model on complex audio and its capacity to follow instructions. Both benchmarks require the model to generate hypotheses directly. We design a unified framework that leverages advanced language models, such as GPT-4, to evaluate the scores of generated hypotheses given the meta-information of the audio. Experimental results demonstrate a high level of consistency between GPT-4-based evaluation and human evaluation. By revealing the limitations of existing LALMs through evaluation results, AIR-Bench can provide insights into the direction of future research. Dataset and evaluation code are available at \href{https://github.com/OFA-Sys/AIR-Bench}{https://github.com/OFA-Sys/AIR-Bench}.
\end{abstract}

\section{Introduction}
Recent advancements in artificial general intelligence have been significantly driven by the emergence of large language models (LLMs)~\citep{gpt3, chatgpt, gpt4, palm, palm2, llama, llama2, qwen}. These models exhibit remarkable abilities in retaining knowledge, engaging in intricate reasoning, and solving problems following human intents. 
Motivated by the striking progress in large language models (LLMs), the domain of large audio-language models (LALMs) has undergone a revolutionary transformation. To perceive and comprehend rich audio signals and further generate textual responses following human instructions, many works have been proposed, such as SALMONN~\citep{2023salmonn}, BLSP~\citep{wang2023blsp}, Speech-LLaMA~\citep{speechllama}, and Qwen-Audio~\citep{qwen-audio}, showcasing promising capabilities for audio-central dialogues.

However, previous LALMs~\citep{2023salmonn, wang2023blsp, speechllama, qwen-audio, AudioGPT, HuggingGPT, LTU, wang2023slm} have predominantly concentrated on evaluation in specific fundamental tasks. The absence of a standardized benchmark for assessing the generative instruction-following abilities of these models has resulted in a reliance on showcasing examples or releasing the chat models for public experimentation to demonstrate their conversational skills. This approach poses significant challenges for conducting fair and objective comparisons across different research endeavors. Moreover, it tends to obscure the models' existing limitations, impeding the ability to monitor advancements within the domain of LALMs. 


For evaluation in audio domains, the majority of research efforts have concentrated on the creation of benchmarks tailored to individual tasks such as LibriSpeech~\citep{panayotov2015librispeech} and Common Voice benchmark~\citep{ardila2019common} for ASR.
Beyond task-specific ones, benchmarks like SUPERB~\citep{SUPERB2021} and HEAR~\citep{HEAR2021} have been designed to test the versatility of self-supervised learning models in a wide variety of tasks. Regarding the assessment of LALMs' ability to follow instructions, to the best of our knowledge, Dynamic-SUPERB~\citep{DynamicSuperb} is the only benchmark devoted to this aspect. Nevertheless, Dynamic-SUPERB only focuses on human speech processing and does not extend to the assessment of models' capabilities in producing open-ended generations such as dialogues.

In this paper, we present AIR-Bench (\textbf{A}udio \textbf{I}nst\textbf{R}uction \textbf{Bench}mark), a novel benchmark designed to evaluate the ability of LALMs to comprehend various audio signals and to interact following instructions. AIR-Bench is characterized by three primary features: 1) \textbf{Comprehensive audio signals coverage}. AIR-Bench offers comprehensive coverage of audio signals, including human speech, natural sounds, and music, ensuring a comprehensive evaluation of LALMs' capabilities. 2) \textbf{Hierarchical Benchmark Structure}. The benchmark consists of \textit{foundation} and \textit{chat} benchmarks. The foundation benchmark comprises 19 distinct audio tasks with over 19,000 single-choice questions, with each question focusing only on a specific foundational ability. GPT-4~\citep{gpt4} extends the questions and candidate choices using dedicated designed prompts. 
The chat component consists of over 2,000 audio-prompted open-ended questions. To enhance the complexity of the audio and achieve a closer resemblance to the intricate audio encountered in real-life situations, we propose a novel audio mixing strategy that incorporates loudness control and temporal dislocation. Specifically, we adjust the loudness and introduce different temporal offsets during the mixing process of two audio clips. The resulting variations in relative loudness and temporal location are then recorded as additional meta-information, contributing to a more comprehensive textual representation of the audio. The quality of data is upheld through automated filtering by GPT-4, followed by manual verification. 3) \textbf{Unified, objective, and reproducible evaluation framework}. Models are required to generate hypothesis sequences directly across both benchmarks to align more accurately with practical scenarios. Then, we employ GPT-4 to generate reference answers given meta-information through carefully constructed prompts. Given references and hypotheses, following~\citet{liu2023mmbench, bai2023touchstone}, we use GPT-4~\citep{gpt4} to judge whether the choice is correct for the foundation benchmark or score hypotheses for the chat benchmark. We further perform 
a second scoring by swapping their positions to eliminate the position bias. Based on comprehensive experiments on 9 LALMs, we observe that existing LALMs either have limited audio understanding or instruction-following capabilities, leaving significant room for improvement in this field.

Our contribution is summarized below:
\begin{itemize}
    \item AIR-Bench is the first generative evaluation benchmark for large audio-language models, encompassing a wide array of audio such as speech, natural sounds, and music.  AIR-Bench is a large and hierarchical benchmark, consisting of the foundation benchmark with 19 audio tasks and over 19k single-choice questions, alongside a chat benchmark with over 2k meticulously curated open-ended audio questions for comprehensive evaluation.
    \item We propose a novel audio mixing strategy with loudness control and temporal dislocation to enhance the complexity of the audio.
    \item A unified, objective, and reproducible evaluation framework has been developed to assess the quality of generative hypotheses.
    \item We conducted a thorough evaluation of 9 models for the purpose of benchmarking. 
\end{itemize}


\section{Related Work}
\paragraph{Benchmarks for Audio Processing.}
Previous studies have primarily focused on evaluating the specific fundamental capabilities of models. In the field of speech processing, automatic speech recognition is one of the most popular tasks, with representative benchmarks including Librispeech~\citep{panayotov2015librispeech}, Common Voice~\citep{ardila2019common}, and FLEURS~\citep{FLEURS2022}. Additionally, there are various benchmarks available for different speech processing tasks such as speech-to-text translation~\citep{wang-etal-2020-covost, wang2020covost, jia2022cvss} and emotion recognition~\citep{cao2014crema,livingstone2018ryerson}. 
In the field of sound processing, several benchmarks have emerged such as Clotho~\citep{drossos2020clotho} and Audiocaps~\citep{kim2019audiocaps} for automatic audio captioning,  
and AVQA~\citep{yang2022avqa} for sound question answering.
In the domain of music processing, numerous datasets are available, including  MusicCaps~\citep{agostinelli2023musiclm} for automatic music captioning, 
and MUSIC-AVQA~\citep{li2022learning} for music question answering. Note that most existing question-answering benchmarks, such as Clotho-AQA, AVQA, and MUSIC-AVQA, have highly constrained answer formats for ease of close-ended evaluation or conversion into classification tasks, rather than supporting open-ended generation. 

Besides the aforementioned datasets that focus on specific tasks, there are benchmarks like SUPERB~\citep{yang2021superb} and HEAR~\citep{turian2022hear} for comprehensive evaluation of self-supervised learning models. When it comes to assessing the ability of LALMs to follow instructions, Dynamic-SUPERB is the only benchmark dedicated to this aspect. However, Dynamic-SUPERB focuses on human speech processing and does not cover open-ended dialogue generation. In contrast, AIR-Bench is the first large-scale generative evaluation benchmark for large audio-language models, encompassing various audio types such as speech, natural sounds, and music.

\paragraph{Large Audio-Language Models following Human Instruction}
Recently, there has been significant interest in instruction-following end-to-end audio-language models. Several models have emerged, each focusing on different audio domains. For instance, there are models specifically focusing on speech processing, such as SpeechGPT~\citep{zhang2023speechgpt}, BLSP~\citep{wang2023blsp}, and LLaSM~\citep{shu2023llasm}. Similarly, there are models tailored for sound processing, like LTU~\citep{LTU}, and for music processing, such as LLark~\citep{gardner2023llark}. In contrast, SALMONN~\citep{tang2023salmonn} and Qwen-Audio~\citep{qwen-audio} are trained using various audio types, showcasing strong universal audio understanding abilities. However, these models are evaluated on different fundamental tasks, making it difficult to conduct a fair comparison. Furthermore, these models rely on showcasing examples or public demos to demonstrate their conversational skills and do not perform rigorous experiments to evaluate their instruction-following abilities. To address these issues, this paper introduces AIR-Bench, which proposes two benchmarks - the foundation benchmark and the chat benchmark, enabling a fair comparison of the models' foundational abilities and their high-level instruction-following capabilities respectively.


\section{AIR-Bench}
%
There exist three unique characteristics that differentiate AIR-Bench from existing benchmarks for
audio understanding: i) AIR-Bench is the first work to incorporate task evaluation from all types of audio in a hierarchical taxonomy; ii) AIR-Bench is the first generative evaluation benchmark that handles the free-form output of LALMs; iii) AIR-Bench adopts GPT-4-based automatic evaluation yielding trustworthy evaluation results with affordable cost. In Sec.~3.1, we present the hierarchical
taxonomy of AIR-Bench and discuss the design philosophy behind it. In Sec.~3.2 and Sec.~3.3, we
introduce how we collect the audio-central question-answer pairs for foundation and chat tasks. In Sec.~3.4, we present the evaluation framework.

\subsection{Overview}
\begin{figure}[t]
  \centering
  \includegraphics[width=0.45\textwidth]{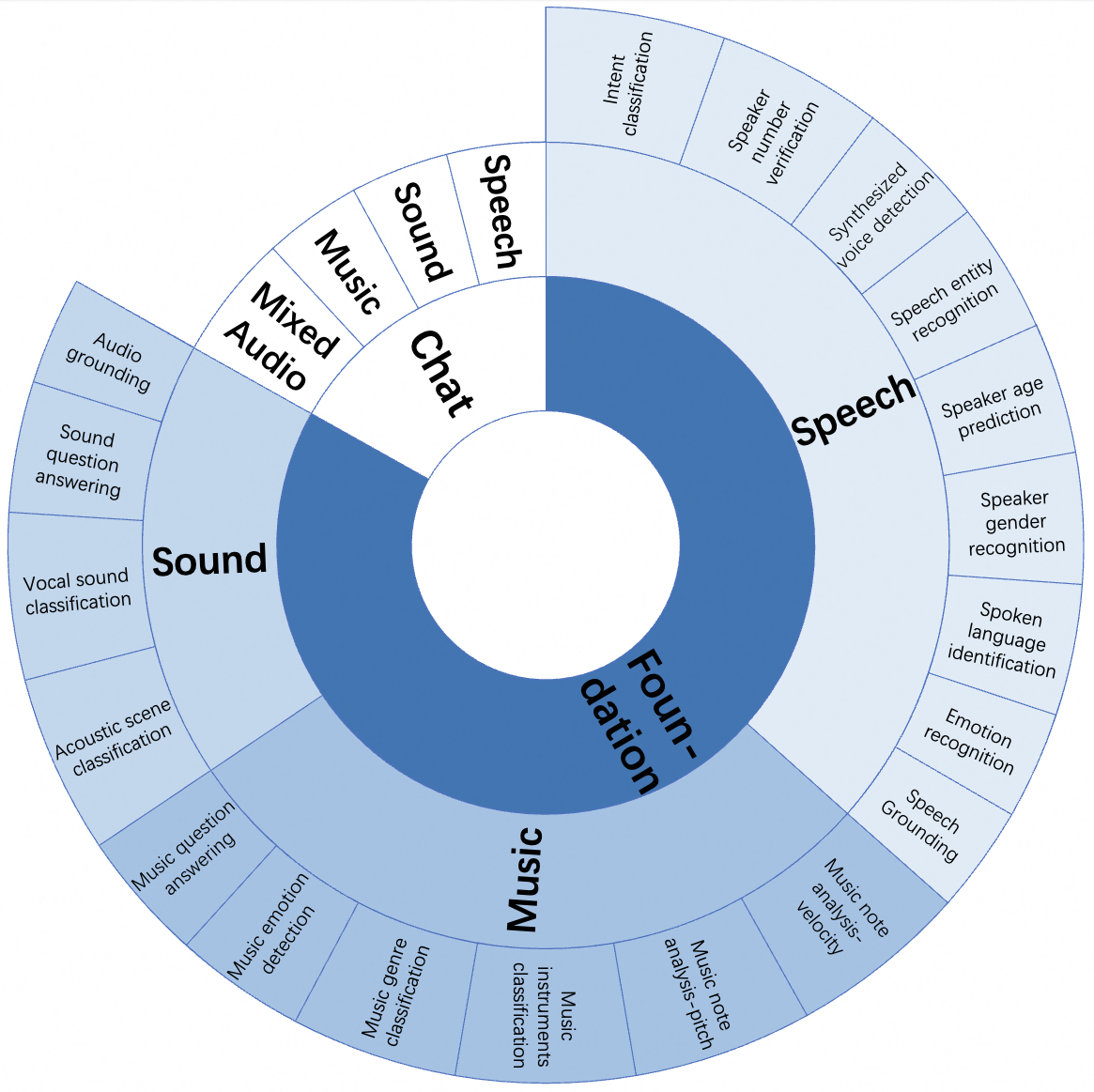}
  \caption{The overview of AIR-Bench. AIR-Bench includes a range of ability dimensions, namely the \textit{foundation} and \textit{chat} abilities, which cater to various audio types such as speech, sound, and music. The foundational dimension comprises 19 distinct leaf abilities, each of which is assessed using a single-choice question format. The chat dimension assesses abilities through an open-ended question-and-answer format, incorporating diverse audio sources and mixed audio.}
  \label{fig:main_figure}
  \vspace{-2mm}
\end{figure}

Chat interaction based on audio is a complex task that encompasses a variety of fundamental competencies. For instance, humans are able to respond to sound events due to their capacities for sound perception and common sense reasoning. Similarly, the ability to respond to others' spoken words is predicated on foundational skills such as speech-to-text recognition and emotion recognition. Based on the motivation, we propose the hierarchical benchmark AIR-Bench by dividing it into \textit{foundation} and \textit{chat} benchmarks. The fundamental one is designed to assess capabilities across individual subtasks, serving to diagnose weaknesses within the model, while the chat benchmark directly evaluates complicated audio-based open-ended questions. The data sample is denoted as $(A,Q,R)$, where $A$ denotes the audio, $Q$ represents the query and $R$ is the reference answer. 
\begin{itemize}
    \item \textbf{Foundation benchmark}: The purpose of the benchmark is to evaluate the individual capabilities of foundational tasks. To reduce the task difficulties and enable the evaluation of various models, we utilize the single-choice question-answering format. Specifically, the query $Q$ is formed by concatenating a question $q$ and candidate choices $C$, denoted as $Q=(q, C)$. We curate a collection of 19 audio tasks that span multiple audio types, such as speech, music, and sound. These tasks include tasks like emotion recognition, acoustic scene classification, and music QA.~\footnote{For transcription tasks such as ASR, we incorporate them into the chat benchmark since they are not suitable for the single-choice task format.}
    \item \textbf{Chat benchmark}: The benchmark encompasses any form of question and answer pairs that could arise from audio signals, with the aim of reflecting the model's ability to genuinely follow user instructions to perform perceiving, reasoning, and interacting within real-world applications. According to the type of audio, the benchmark is categorized into four dimensions: speech, sound, music, and mixed audio, where mixed audio refers to audio that is a mixture of multiple types of audio, such as human voice with background music.
\end{itemize}

The overview of AIR-Bench is shown in Fig.~\ref{fig:main_figure}. 

\subsection{Foundation Benchmark}

\begin{table}[t]
\centering
\resizebox{0.95\columnwidth}{!}{%
\begin{tabular}{cccc}
\toprule
\textbf{Types} & \textbf{Task} & \textbf{Dataset-Source} & \textbf{Num} \\ \hline
\multirow{9}{*}{\textbf{Speech}} & Speech grounding & Librispeech~\citep{panayotov2015librispeech} & 0.9k \\ \cline{2-4} 
 & Spoken language identification & Covost2~\citep{wang2020covost} & 1k \\ \cline{2-4} 
 & \begin{tabular}[c]{@{}c@{}}Speaker gender recognition\\ (biologically)\end{tabular} & \begin{tabular}[c]{@{}c@{}}Common voice~\citep{ardila2019common} \\ MELD~\citep{poria2018meld}\end{tabular} & 1k \\ \cline{2-4} 
 & Emotion recognition & \begin{tabular}[c]{@{}c@{}}IEMOCAP~\citep{busso2008iemocap} \\ MELD~\citep{poria2018meld}\end{tabular} & 1k \\ \cline{2-4} 
 & Speaker age prediction & Common voice~\citep{ardila2019common} & 1k \\ \cline{2-4} 
 & Speech entity recognition & SLURP~\citep{bastianelli2020slurp} & 1k \\ \cline{2-4} 
 & Intent classification & SLURP~\citep{bastianelli2020slurp} & 1k \\ \cline{2-4} 
 & Speaker number verification & VoxCeleb1~\citep{nagrani2020voxceleb} & 1k \\ \cline{2-4} 
 & Synthesized voice detection & FoR~\citep{reimao2019dataset} & 1k \\ \hline
\multirow{4}{*}{\textbf{Sound}} & Audio grounding & AudioGrounding~\citep{xu2021text} & 0.9k \\ \cline{2-4} 
 & Vocal sound classification & VocalSound~\citep{gong2022vocalsound} & 1k \\ \cline{2-4} 
 & Acoustic scene classification & \begin{tabular}[c]{@{}c@{}}CochlScene~\citep{jeong2022cochlscene} \\ TUT2017~\citep{mesaros2017dcase}\end{tabular} & 1k \\ \cline{2-4} 
 & Sound question answering & \begin{tabular}[c]{@{}c@{}}Clotho-AQA~\citep{lipping2022clotho} \\ AVQA~\citep{yang2022avqa}\end{tabular} & 1k \\ \hline
\multirow{7}{*}{\textbf{Music}} & Music instruments classification & \begin{tabular}[c]{@{}c@{}}Nsynth~\citep{engel2017neural}\\ MTJ-Jamendo~\citep{bogdanov2019mtg}\end{tabular} & 1k \\ \cline{2-4} 
 & Music genre classification & \begin{tabular}[c]{@{}c@{}}FMA~\citep{defferrard2016fma} \\ MTJ-Jamendo~\citep{bogdanov2019mtg}\end{tabular} & 1k \\ \cline{2-4} 
 & Music note analysis-pitch & Nsynth~\citep{engel2017neural} & 1k \\ \cline{2-4} 
 & Music note analysis-velocity & Nsynth~\citep{engel2017neural} & 1k \\ \cline{2-4} 
 & Music question answering & MUSIC-AVQA~\citep{li2022learning} & 0.8k \\ \cline{2-4} 
 \cline{2-4} 
 & Music emotion detection & MTJ-Jamendo~\citep{bogdanov2019mtg} & 1k \\
 \bottomrule
\end{tabular}%
}
\caption{The statistics of the foundation benchmark.}
\label{table_fundamental}
\vspace{-2mm}
\end{table}

\begin{table}[]
\resizebox{\columnwidth}{!}{%
\begin{tabular}{cccl}
\toprule
\textbf{Types} & \textbf{Dataset-Source} & \textbf{Num} & \multicolumn{1}{c}{\textbf{Question Example}} \\ \hline
\textbf{Speech} & \begin{tabular}[c]{@{}c@{}}Fisher~\citep{cieri2004fisher} \\ SpokenWOZ~\citep{si2023spokenwoz}\\ IEMOCAP~\citep{busso2008iemocap}\\  Common voice~\citep{ardila2019common}\end{tabular} & 800 & \begin{tabular}[c]{@{}l@{}}Did the first speaker have any more\\ questions or need further information?\end{tabular} \\ \hline
\textbf{Sound} & Clotho~\citep{drossos2020clotho} & 400 & \begin{tabular}[c]{@{}l@{}}What should you do to the cloth \\ according to the voice in the audio?\end{tabular} \\ \hline
\textbf{Music} & MusicCaps~\citep{agostinelli2023musiclm} & 400 & \begin{tabular}[c]{@{}l@{}}How might the elements of the music\\ in the audio, despite its poor sound\\ quality, musically convey a sense of\\ patriotism and ceremonial grandeur\\ within a 150-word essay?\end{tabular} \\ \hline
\multirow{2}{*}{\textbf{\begin{tabular}[c]{@{}c@{}}Mixed \\ \\ Audio\end{tabular}}} & \begin{tabular}[c]{@{}c@{}}Common voice~\citep{ardila2019common}\\ AudioCaps~\citep{audiocaps}\end{tabular} & 200 & \begin{tabular}[c]{@{}l@{}}What sound is heard along with the male\\ speaker in his twenties?\end{tabular} \\ \cline{2-4} 
 & \begin{tabular}[c]{@{}c@{}}Common voice~\citep{ardila2019common}\\ MusicCaps~\citep{agostinelli2023musiclm}\end{tabular} & 200 & \begin{tabular}[c]{@{}l@{}}What type of melody can be heard in the\\ background of the male speaker's audio?\end{tabular} \\ 
\bottomrule
\end{tabular}%
}
\caption{The statistics and examples of the chat benchmark.}
\label{table_chat}
\end{table}

\paragraph{Data Source.}
We collected over 19k data samples for the foundation dimension, encompassing 19 different subtasks. The data source and statistics are provided in Table~\ref{table_fundamental}. To ensure a fair and comprehensive evaluation of each capability, we aimed for an even distribution of problems related to different abilities during the data collection process. All audio sources were obtained from the original dev or test subsets to prevent data leakage.

\paragraph{Single-choice Query and Reference.} The query $Q$ is formed by concatenating a question $q$ and candidate choices $C$. For the question $q$, we mainly construct questions through GPT-4~\citep{gpt4}, except for QA tasks since the datasets inherently contain questions and we can directly re-use them. Specifically, we design the prompt for the distinct task and provide three questions as demonstrations. Subsequently, GPT-4 generates additional diverse questions based on these inputs. The generated questions are manually reviewed, and 50 different questions are selected for each task. The variability in question format aims to evaluate the model's ability to follow instructions rather than being overly reliant on specific templates. For each question, we further generate candidate choices $C$ from different sources: 1) For tasks with choices in original datasets like AVQA~\citep{yang2022avqa}, we directly re-use it; 2) For classification tasks, we randomly select options from the predetermined set of categories to serve as candidate choices;
3) For other tasks, we prompt GPT-4 to generate candidate choices directly, consisting of one correct option and three incorrect options. We encourage these incorrect options to resemble the correct one, making the single-choice task more challenging. The reference answer is the golden correct choice. To avoid position bias, the candidate choices are randomly shuffled. We provide examples of each task in Table~\ref{Table_Examples_of_Foundation} of the Appendix.


\subsection{Chat Benchmark}
\begin{figure}[h!]
  \centering
  \includegraphics[width=0.45\textwidth]{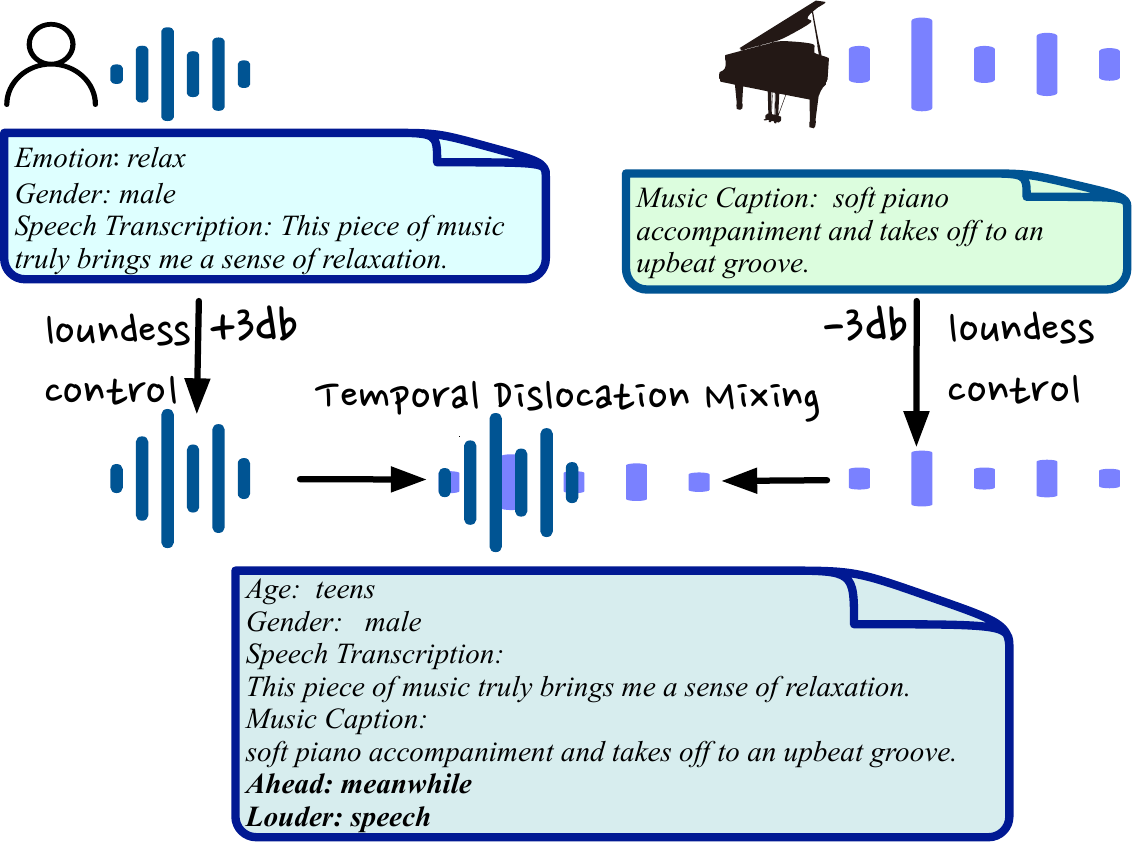}
  \caption{Loudness and temporal location controlled mixing strategy. Loudness control aims to provide \textit{Louder} meta-information, indicating which audio clip exhibits a higher volume. Temporal dislocation mixing aims to provide the \textit{Ahead} meta-information, referring to the temporal relationship between the two audio clips.}
  \label{fig_mixing_strategy}
\end{figure}

\begin{figure*}[t!]
  \centering
  \vspace{-1mm}
  \includegraphics[width=0.95\textwidth]{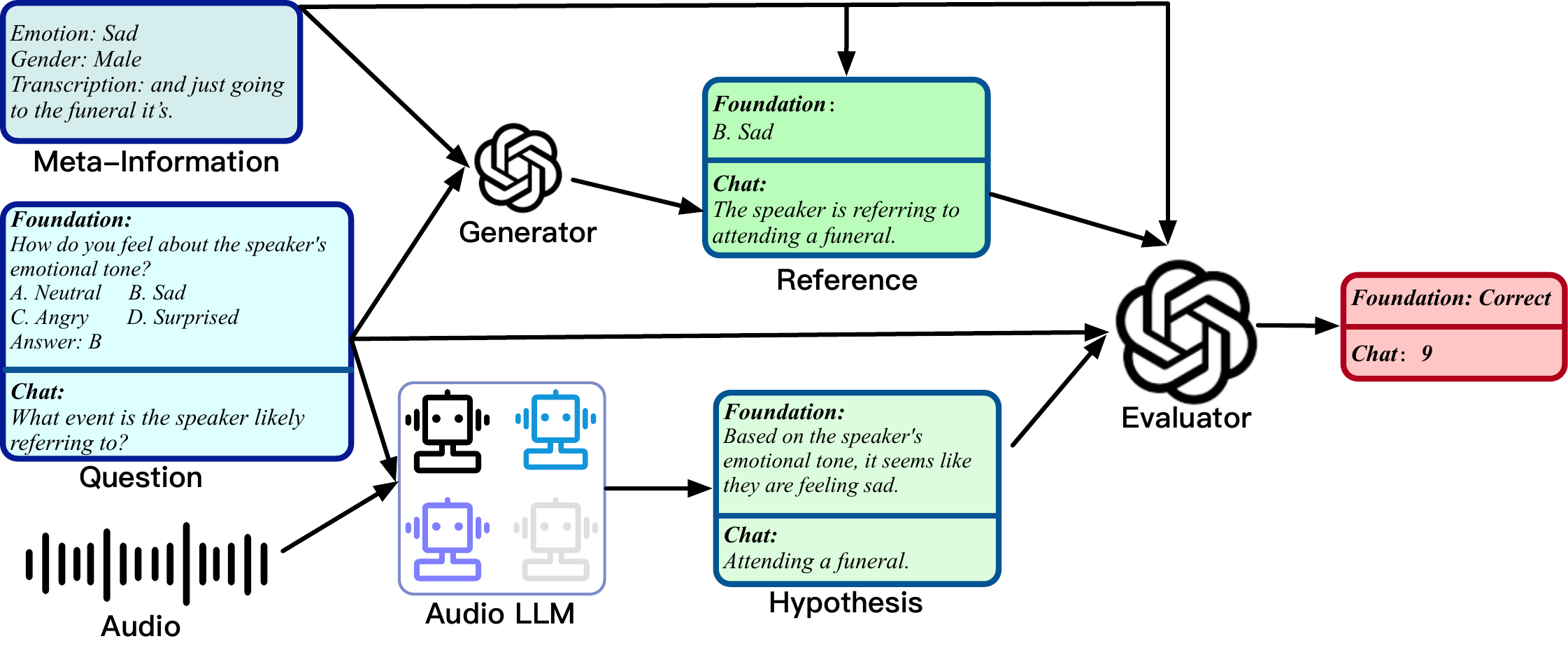}
  \caption{Automated generative evaluation for large audio-language models (LALMs). In the evaluation framework, LALMs are provided with audio input along with a corresponding question, following which they generate a hypothesis. The performance of the hypothesis is then assessed using the GPT evaluator, which compares it against a reference answer by considering the meta-information and the question. For the foundation benchmark, the reference answer is the golden choice extracted from the meta-information, and the evaluation score is binary, with 0 indicating an incorrect answer and 1 representing a correct answer. For the chat benchmark, the reference answer is produced by the GPT-4 generator. The reference answer serves as a reference for scoring, stabilizing the scoring process. The output score for the chat benchmark ranges from 1 to 10, based on the assessment of usefulness, relevance, accuracy, and comprehensiveness of the hypothesis.}
  \label{fig_Eval_Framework}
\end{figure*}

\paragraph{Data Source and Audio Mixing Strategy.} As shown in Table~\ref{table_chat}, we have collected more than 2k data samples spanning various audio types including speech, sound, music, and mixed audio. The purpose of introducing mixed audio is to augment the complexity of the audio signals and make it closer to audio from real-world audio scenarios. To achieve this, we propose a novel mixing strategy involving loudness control and temporal dislocation, as illustrated in Fig.~\ref{fig_mixing_strategy}. Specifically, we can adjust the relative loudness and temporal relationship between two audio clips for mixing. Then, we can create a complex audio signal that combines their meta-information, such as speech transcription accompanied by a background music caption. Furthermore, the meta-information also includes labels indicating which audio clip is louder and which is ahead in the temporal sequence.

\paragraph{Open-ended Query and Reference.}
To prompt GPT-4 to generate open-ended question-answer pairs for audio, we should interpret the rich information in each audio with texts. We collect all of \textit{meta-information} such as gender, age, emotion, transcription, language for speech, caption for natural sound, and instrument, caption for music from the original dataset. Rather than relying on pre-trained models to extract this meta-information for each audio clip, we adopt the ground truth meta-information to avoid potential errors.

After gathering meta-information about the audio, we manually construct prompts (see Appendix~\ref{fig_GPT_Prompts} for guiding GPT-4 in generating question-answer pairs that specifically focus on different abilities). These prompts are carefully designed to ensure a comprehensive coverage of chat interactions, taking into consideration the diverse range of audio signals involved. We design the prompts to facilitate the generation of questions related to the perception and reasoning for different types of audio. For the natural sound, the prompts are further tailored to generate questions that involve determining appropriate responses to sound events within a specific scenario. For the music category, prompts are devised to elicit creative writing and story-generation questions based on music composition. To ensure the quality of the generated results, these prompts are designed in a manner that enables GPT-4 to automatically filter out responses that are not directly related to audio. Additionally, we manually reviewed all the question-answer pairs to ensure the quality of the questions and the reliability of the answers. The generated answers from GPT-4 are considered as references.

\subsection{Evaluation Strategy}
In this paper, we leverage a unified evaluation method, as shown in Fig.~\ref{fig_Eval_Framework}, by viewing both the single-choice question in the foundation benchmark, and the open-ended question in the chat benchmark, as the generation tasks for the purpose of better alignment with actual use case scenarios of LALMs. That is, given audio and questions, LALMs are required to directly generate the answers as hypotheses, rather than comparing the perplexity on the probability of different reference answers via teacher forcing. Automated and accurate evaluation of open-ended generation is a challenging problem. Traditional automatic metrics such as WER, ROUGE~\citep{lin2004rouge}, METEOR~\citep{BanerjeeL05} have shown a low correlation with human judgments~\citep{liu2023gpteval}. Recently, LLM-based evaluation, such as GPT-4, shows better human preference alignment~\citep{zheng2023judging, liu2023gpteval}. In this work, we adopt reference-based GPT-4 evaluators to judge the generation quality of LALMs in the audio domain.


However, GPT-4 cannot be directly used as an evaluator since it cannot receive audio inputs. To address this limitation, we 
offer the GPT-4 model rich meta-information of audio to replace audio input. Subsequently, we present questions and employ GPT-4 to evaluate the hypotheses produced by LALMs. To ensure consistency and fairness for evaluation, each model's answer is compared against the same reference answer for scoring. For the foundation benchmark, the reference answer is the golden choice, and we prompt the evaluator to determine whether the hypothesis is correct or not. For the chat benchmark, the reference answer is generated by GPT-4, and we prompt the evaluator to provide a score ranging from 1 to 10, based on the assessment of usefulness, relevance, accuracy, and comprehensiveness of the hypothesis. The prompts used in the evaluation process can be found in Appendix~\ref{fig_GPT_Prompts}. Note that for the chat benchmark, the role of the reference is not to serve as the ground truth answer, but rather as a reference for scoring by GPT-4, in order to stabilize its scoring. Additionally, to mitigate any potential position bias resulting from the order of hypothesis and reference, following~\citet{bai2023touchstone}, we perform a second scoring round by swapping their positions and then compute the average of the two scores. Unless otherwise specified, the GPT-4 evaluator is GPT-4 Turbo, the \textit{gpt-4-0125-preview} version~\footnote{https://platform.openai.com/docs/models/gpt-4-and-gpt-4-turbo}.

\begin{table*}[t]
\resizebox{\textwidth}{!}{%
\begin{tabular}{c|cccclccccc}
\toprule
\textbf{Benchmark} & \multicolumn{4}{c}{\textbf{Foundation}} &  & \multicolumn{5}{c}{\textbf{Chat}} \\ \hline
\textbf{Categories} & \textbf{Speech} & \textbf{Sound} & \textbf{Music} & \textbf{Average} &  & \textbf{Speech} & \textbf{Sound} & \textbf{Music} & \textbf{\begin{tabular}[c]{@{}c@{}}Mixed\\ Audio\end{tabular}} & \textbf{Average} \\ \hline
\textbf{SALMONN} & 37.8\% & 33.0\% & 37.1\% & 36.0\% &  & 6.16 & 6.28 & 5.95 & \textbf{6.08} & 6.11 \\ \hline
\textbf{Qwen-Audio-Chat} & 58.7\% & 60.2\% & 44.8\% & 54.5\% &  & 6.47 & \textbf{6.95} & 5.52 & 5.38 & 6.08 \\ \hline
\textbf{Qwen-Audio Turbo} & \textbf{63.4}\% & \textbf{61.0\%} & \textbf{48.9\%} & \textbf{57.8\%} &  & 7.04 & 6.59 & \textbf{5.98} & 5.77 & \textbf{6.34} \\ \hline
\textbf{BLSP} & 36.6\% & 31.4\% & 26.1\% & 31.4\% &  & 6.17 & 5.55 & 5.08 & 4.52 & 5.33 \\ \hline
\textbf{PandaGPT} & 39.0\% & 43.6\% & 38.1\% & 40.2\% &  & 3.58 & 5.46 & 5.06 & 2.93 & 4.25 \\ \hline
\textbf{Macaw-LLM} & 32.2\% & 30.1\% & 29.7\% & 30.7\% &  & 0.97 & 1.01 & 0.91 & 1.00 & 1.01 \\ \hline
\textbf{SpeechGPT} & 34.3\% & 27.5\% & 28.1\% & 30.0\% &  & 1.57 & 0.95 & 0.95 & 1.14 & 1.15 \\ \hline
\textbf{NExT-GPT} & 33.6\% & 32.2\% & 28.9\% & 31.5\% &  & 3.86 & 4.76 & 4.18 & 2.92 & 4.13 \\ \hline
\textbf{Whisper+GPT-4} & 53.6\% & / & / & / &  & \textbf{7.54} & / & / & / & / \\ 
\bottomrule
\end{tabular}%
}
\caption{The comparison of different LALMs on AIR-Bench.}
\label{table_main_results}
\end{table*}

\section{Experiments}

\begin{table}[h]
\centering
\vspace{-2mm}
\resizebox{0.85\columnwidth}{!}{%
\begin{tabular}{c|cc}
\toprule
\textbf{Model Name} & \textbf{\begin{tabular}[c]{@{}c@{}}Exact \\Matching\end{tabular}} & \textbf{\begin{tabular}[c]{@{}c@{}}GPT\\ Align\end{tabular}} \\ \hline
\textbf{SALMONN} & 97.3\% & 100.0\% \\ \hline
\textbf{Qwen-Audio-Chat} & 30.7\% & 100.0\% \\ \hline
\textbf{Qwen-Audio Turbo} & 48.2\% & 100.0\% \\ \hline
\textbf{BLSP} & 100.0\% & 100.0\% \\ \hline
\textbf{PandaGPT} & 30.8\% & 100.0\% \\ \hline
\textbf{Macaw-LLM} & 0.1\% & 100.0\% \\ \hline
\textbf{SpeechGPT} & 0.0\% & 100.0\% \\ \hline
\textbf{NExT-GPT} & 98.1\% & 100.0\% \\ 
\bottomrule
\end{tabular}%
}
\caption{The success rate of different strategies of matching hypotheses with the golden choices for the foundation benchmark. The success rate denotes the probability that the model successfully responds to one of the choices.}
\vspace{-2mm}
\label{Table_success_rate}
\end{table}

\subsection{Models}
We evaluate the performance of various LALMs with instruction-following capabilities. These models are either open-sourced or accessible through public APIs, such as SpeechGPT~\citep{zhang2023speechgpt}, BLSP~\citep{wang2023blsp}, SALMONN~\citep{2023salmonn}, Qwen-Audio-Chat~\citep{qwen-audio}, and Qwen-Audio Turbo~\footnote{https://help.aliyun.com/zh/dashscope/developer-reference/qwen-audio-api}. Additionally, we consider large multi-modality models with audio understanding abilities like PandaGPT~\citep{su2023pandagpt}, Macaw-LLM~\citep{Macaw-LLM}, and NExT-GPT~\citep{NextGPT}. Besides, we also incorporate a sequential approach comprising Whisper-large-v2~\citep{radford2023robust} and GPT-4 Turbo~\citep{gpt4} for tasks related to speech as a baseline. We evaluate the performance of all these models on both fundamental and chat benchmarks, utilizing their latest publicly available checkpoints. In cases of multiple checkpoints, we select the model with the largest parameter size. For all models, we directly follow their default decoding strategies for evaluation.

\begin{figure*}[t!]
  \centering
  \includegraphics[width=1.0\textwidth]{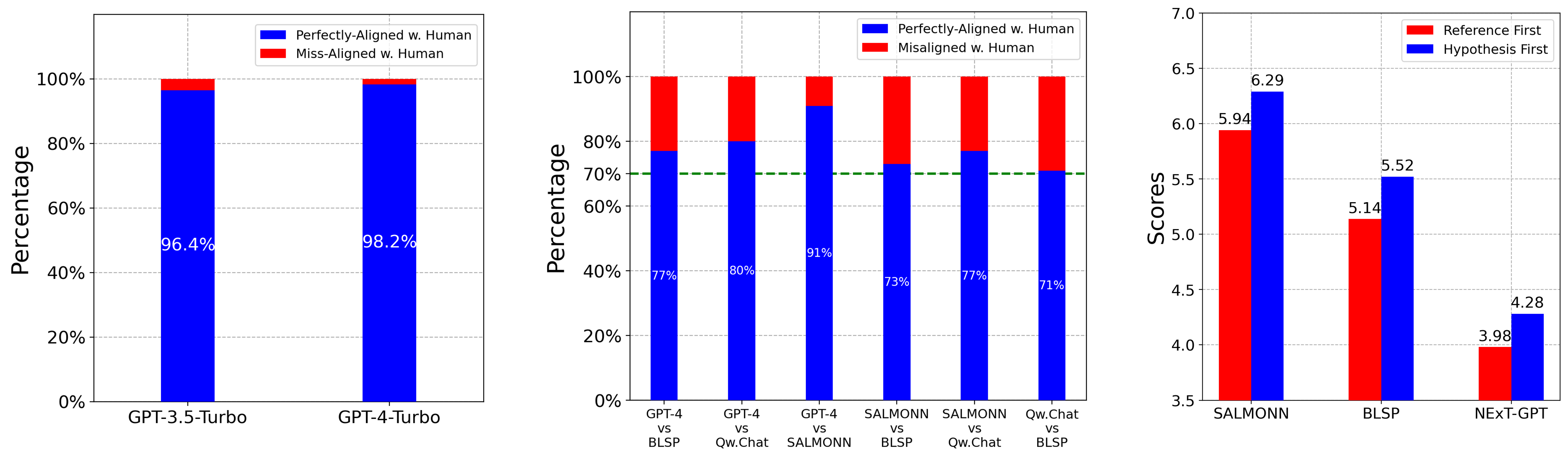}
  \vspace{-1mm}
  \renewcommand{\arraystretch}{0.4} 
  \begin{tabular}{lll}
  \hspace{0.35cm}\small(a)   &
  \hspace{1.8cm} \small(b) Human evaluation  for   & 
  \hspace{2.0cm}\small(c) Positional bias of \\
  \hspace{0.8cm}\small the foundation benchmark  &
  \hspace{2.35cm}\small the chat benchmark  & 
  \hspace{2.45cm}\small evaluation \\
  
  \end{tabular}
  \vspace{-1mm}
  \caption{The experiments of human evaluation and the position bias of GPT-4 evaluator. Figure~(a) and (b) are the results of consistency between the GPT-4 evaluator and human judgment on the foundation benchmark and chat benchmark, respectively. Figure~(c) refers to the result of scores by interchanging the position of the hypothesis and reference during evaluation on the chat benchmark.}
  \label{fig_three_figures}
\end{figure*}

\subsection{Main Results}

The results of LALMs are presented in Table~\ref{table_main_results}. The detailed results are shown in Table~\ref{Table_Foundation_Results}. For the foundation benchmark, we also conduct a comparison between the use of an exact matching strategy with our proposed GPT-4 alignment strategy. As an example, we try to match `B', `B.', `B)', \textit{etc.} with LALMs' hypothesis for the exact matching. The results are shown in Table~\ref{Table_success_rate}. We can find that BLSP and SALMONN have a high success rate in directly generating the choice, showcasing their strong ability to follow single-choice instruction. However, we find that it is challenging to precisely extract the predicted choice from the hypotheses of other models due to significant variations in the output formats of different LALMs. However, with the assistance of GPT-4 as the evaluator, the success rate for all models can be improved to 100\%.

According to Table~\ref{table_main_results}, Qwen-Audio-Chat and Qwen-Audio Turbo demonstrate superior performance in the foundation benchmark, surpassing other models in the domains of speech, sound, and music. Second to the two models, PandaGPT and SALMONN also exhibit noteworthy performances. Regarding the chat benchmark, Qwen-Audio Turbo achieves the highest average score, followed by SALMONN and Qwen-Audio-Chat with scores of 6.11 and 6.08, respectively.
Among these models, SALMONN outperforms others in terms of mixed audio understanding. Note that the speech dimension in the foundation benchmark includes tasks beyond speech transcriptions, such as speaker gender, age, and emotion prediction, while the speech in the chat benchmark primarily revolves around speech transcriptions. Thus, Whisper plus GPT-4 receives a relatively low score in the foundation benchmark but obtains the highest score in the chat benchmark.

Based on these results, we have several observations: 1) The existing LALMs either have limited audio understanding or instruction-following capabilities. For instance, Qwen-Audio Turbo achieves the highest average score in both foundation and chat benchmarks while the model displays a weak proficiency in following single-choice instructions such as often directly generating a full sentence semantically akin to one of the choices, and thus receives a low success rate for the exact matching; 2) As for chat abilities related only to speech transcription, none of the models surpass the sequential baseline Whisper plus GPT-4.

\subsection{Human Evaluation}
\label{sec:human_evaluation}
To evaluate the consistency between the evaluations of GPT-4 and human judgments, we design experiments for both the foundation and chat benchmarks. For the foundation benchmark, we instruct the testers to determine which option aligns closest with the hypothesis. We then compare the option selected by human testers with the option chosen by GPT-4 to assess the extent of agreement. For this consistency analysis, we employed Qwen-Audio-Chat as a representative model and randomly selected 400 questions from the benchmark. These questions were then evaluated by three native English speakers. Additionally, we also compared the performance of GPT-4 with GPT-3.5 Turbo. As depicted in Figure~\ref{fig_three_figures}~(a), GPT-4 Turbo, serving as the evaluator, exhibited a high level of consistency at 98.2\% with human judgments. Comparatively, GPT-3.5 Turbo had a slightly lower consistency rate of 96.4\%.

Regarding the chat benchmark, obtaining a numerical score on a scale of 1 to 10 directly from testers poses challenges. Therefore, we resort to a pairwise comparison of the models instead. Testers listen to audio and compare the performance of both models based on their usefulness, relevance, accuracy, and comprehensiveness to the given question, indicating their preference as either ``A is better'', ``B is better'', or ``Both are equal''.
Subsequently, we convert the GPT-4 scores into the same preference-based rating as the human testers for any two models. We then assess the consistency between the two sets of results. For the chat benchmark, we conduct pairwise comparisons among Qwen-Audio-Chat, SALMONN, BLSP, and GPT-4. We randomly select 200 questions and have them evaluated by three native English speakers. As depicted in Figure~\ref{fig_three_figures}~(b), the pairwise preference consistency scored above 70\%, demonstrating a high level of agreement.


\subsection{Ablation Study of Positional Bias}
In our evaluation framework, we adopt a strategy of scoring twice by interchanging the positions of the hypothesis and reference and calculating the average of the two scores. This approach helps mitigate the bias that may arise from the positional placement. The outcomes of these two evaluations are presented in Figure~\ref{fig_three_figures}~(c). We observe that the GPT-4 evaluator exhibits a clear bias in scoring when the hypothesis is placed before the reference. This highlights the importance of conducting a second scoring to account for addressing this bias.

\section{Conclusion}
In this paper, we present AIR-Bench, the first generative evaluation benchmark designed specifically for audio-language models. AIR-Bench comprises 19 audio tasks with over 19k single-choice questions in the foundation benchmark, as well as over 2k open-ended audio questions in the chat benchmark. Notably, the benchmark covers diverse audio types such as speech, natural sounds, and music. We also propose a novel audio mixing strategy to simulate audio from real-world scenarios more accurately.
A standardized, objective, and reproducible evaluation framework is employed to automatically assess the quality of hypotheses generated by LALMs. We conduct a thorough evaluation of 9 prominent open-source LALMs. Additionally, we plan to launch and maintain a leaderboard that will serve as a platform for the community to access and compare model performance consistently over time.

\section{Limitations}
The objective of AIR-Bench is to develop a large-scale, extensive and generative evaluation framework that encompasses a wide range of audio domains and tasks. However, AIR-Bench currently has several limitations. Firstly, it does not incorporate tasks involving multiple audio comparisons, such as assessing music coherence, for both the foundation and chat benchmark. Besides, AIR-Bench does not encompass the evaluation of multi-turn dialogues that may involve multiple audio inputs. For evaluation, AIR-Bench relies on a powerful and robust evaluator such as GPT-4. However, the availability and accessibility of the GPT-4 API are external factors beyond our control. In the event that GPT-4 transitions to a closed-source model or implements higher pricing standards in the future, alternative evaluators will need to be explored and considered.

\section{Ethical Considerations}
The AIR-Bench initiative uses publicly available datasets to create a collection of relevant question-and-answer data. It then uses automated methods to evaluate this data, which is a more efficient alternative to manually evaluating it. However, there are challenges with this automated evaluation approach, including the potential for data misuse and the introduction of biases. To prevent data misuse, we follow the licenses and usage guidelines associated with the original open-source materials when generating related data. It's important to point out that the automated evaluation could be biased. These biases may come from the datasets themselves or the scoring algorithms used, causing differences between automated evaluation results and human judgment. Therefore, the outcomes obtained from automated evaluations should be viewed with caution and used as a general benchmark, rather than a definitive measure. 

\bibliography{custom}

\appendix
\section{Detailed Results of Foundation Benchmark}
In Table~\ref{Table_Foundation_Results}, we delineate the performance assessment for each model across the various tasks on the foundation benchmark. With the exception of Speaker Gender Recognition and Synthesized Voice Detection, which are binary-choice tasks, all other tasks necessitate a selection from four options. As such, a random selection in the Speaker Gender Recognition and Synthesized Voice Detection datasets would theoretically achieve an accuracy of 50\%, while the expected accuracy for random choices across the remaining datasets stands at 25\%. Consequently, any performance metrics that approximate these random baselines are indicative of an absence of discernible proficiency in the respective tasks.

\section{GPT Prompts for the Chat benchmark}
\begin{figure*}[p] 
  \centering
  {\includegraphics[width=\textwidth,height=\textheight,keepaspectratio]{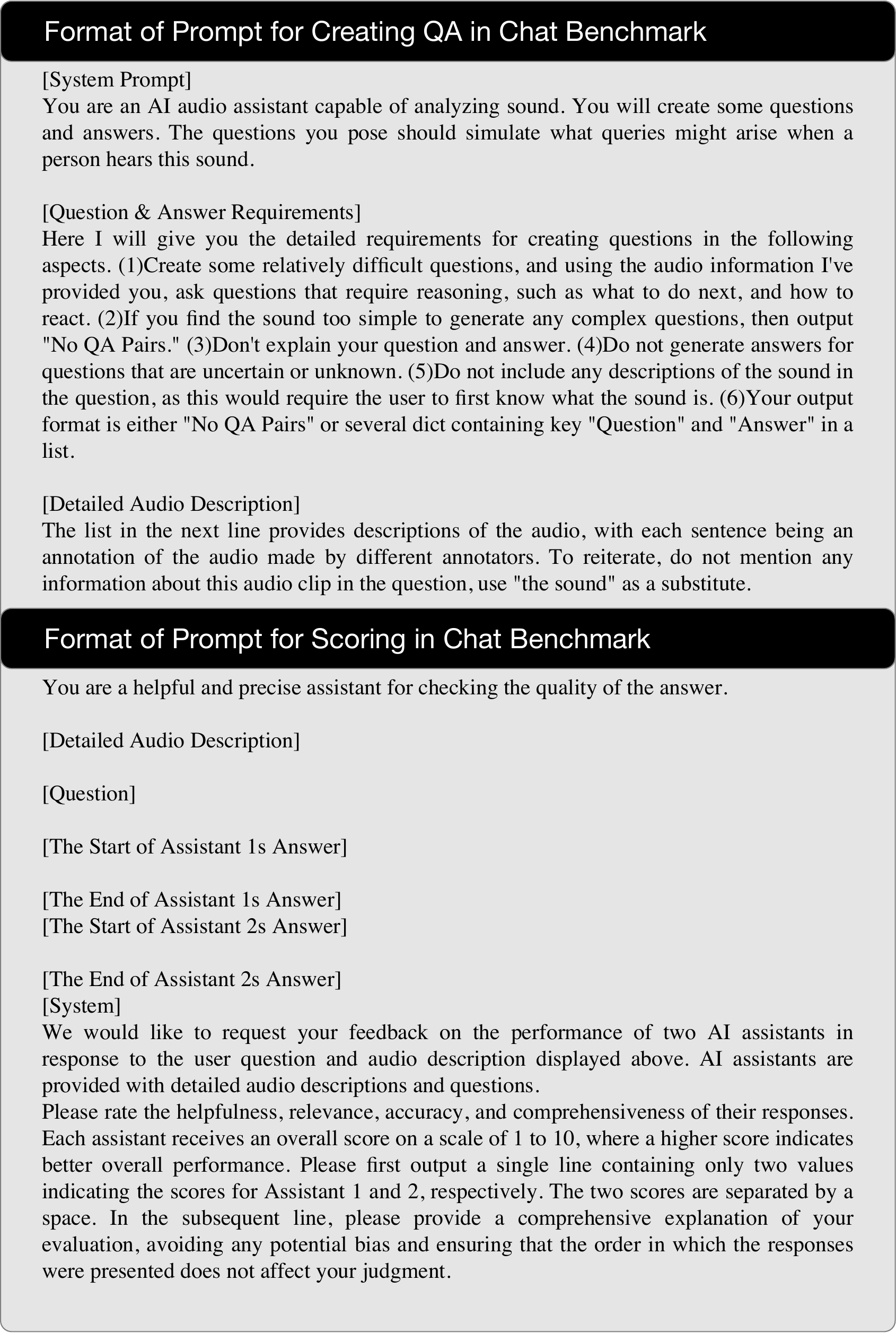} 
  }
  \caption{GPT prompts for creating QA in the foundation benchmark and scoring in the chat benchmark.}
  \label{fig_GPT_Prompts}

\end{figure*}

In Figure~\ref{fig_GPT_Prompts}, we display the carefully crafted prompts that we have developed on our chat benchmark. The figure is divided into two sections, the upper section contains prompts designed specifically for generating question-answer pairs related to reasoning, while the lower section features prompts aimed at assessing the chat performance scores of the models.

When generating questions and reference answers, we guide the process by specifying the type of questions to be elicited, allowing GPT-4 to automatically exclude data that is less amenable to question formulation. For the evaluation of the chat performance scores, we instruct GPT-4 to take a multifaceted approach, scoring both the reference answers and the model responses. This ensures that the reference answers consistently serve as a standard for comparison. 

\section{Prompts Engineering for GPT Scoring}
In this section, we partially demonstrate the process of adjusting the prompt aimed at assessing the chat performance
scores of the models.
\begin{itemize}
    \item If we streamline our prompt by removing the descriptions pertaining to helpfulness, relevance, accuracy, and comprehensiveness, specifically by omitting "Please rate the helpfulness, relevance, accuracy, and comprehensiveness of their responses." and "In the subsequent line, please provide a comprehensive explanation of your evaluation, avoiding any potential bias and ensuring that the order in which the responses were presented does not affect your judgment.", we found that across multiple tests, many responses that were originally scored a perfect 10 were downgraded to a 9 or 8, while the unequivocally incorrect responses saw their scores rise from an initial 1 to a 2 or 3. This suggests that including these 'superfluous' descriptions aids the model in assigning more precise scores during the evaluation process and helps to avoid 'normalization' of scores.
    \item If we change the positioning, such as moving the entire [Detailed Audio Description] section behind the [Question] and [Answer], or swapping the positions of [Question] and [Answer]; these alterations impact the scoring, turning originally correct evaluations incorrect. Absolutely correct answers were inexplicably awarded scores as low as 5, whereas absolutely incorrect responses occasionally received scores around 5 as well. Therefore, our conclusion is that the prompt exhibits a strong sensitivity to the permutation of positions.
Minor punctuation or grammatical errors do not affect the scoring.
\end{itemize}

\section{Examples of the Foundation Benchmark}
In Table~\ref{Table_Examples_of_Foundation}, we present data examples for each task within the foundation benchmark. 

\section{Examples of LALMs' responses}
\begin{figure*}[p] 
  \centering
  \scalebox{0.95}{\includegraphics[width=\textwidth,height=\textheight,keepaspectratio]{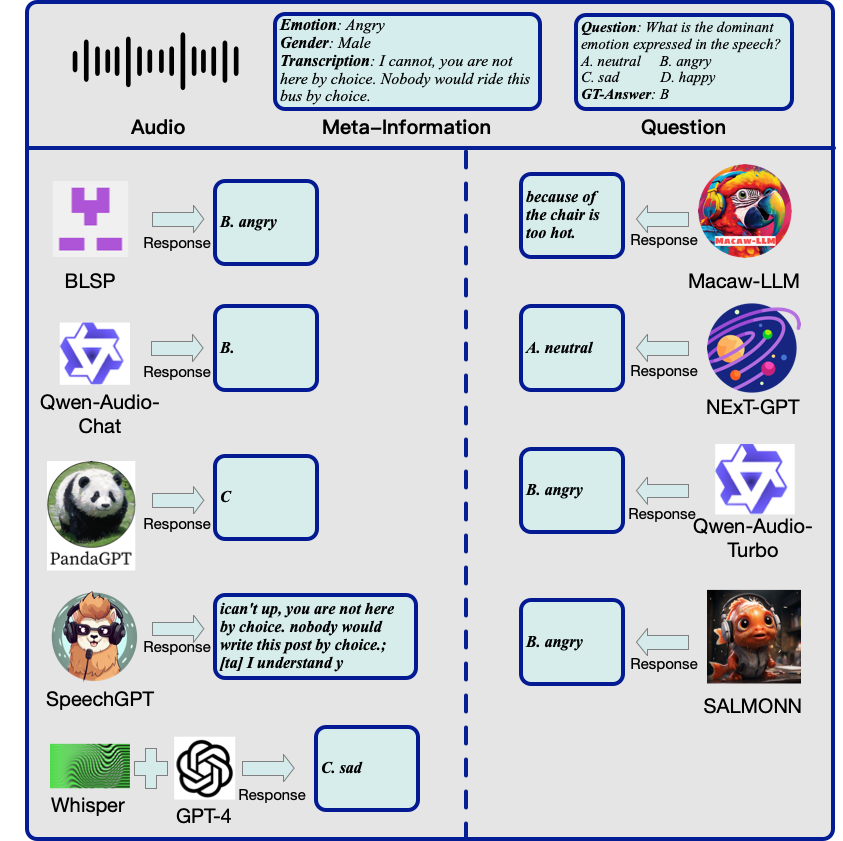} 
  }
  \caption{The illustration of the models' responses on the foundation benchmark.}
  \label{fig_Model_Responses_of_Foundation_Benchmark}
\end{figure*}

\begin{figure*}[p] 
  \centering
  \includegraphics[width=\textwidth,height=\textheight,keepaspectratio]{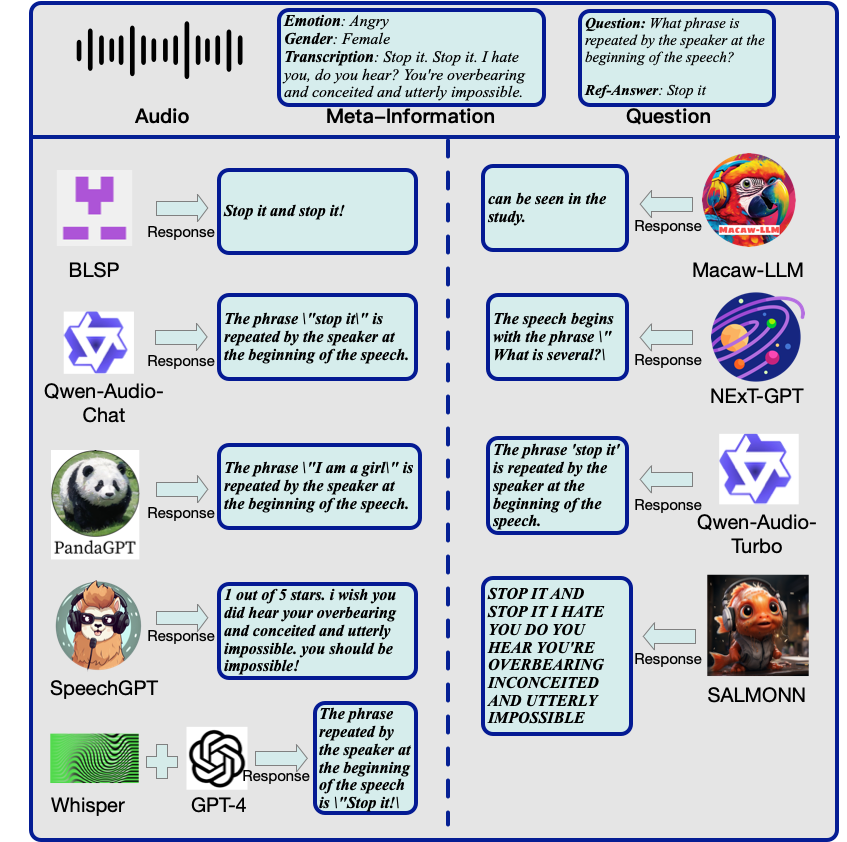} 
  \caption{The illustration of the model's responses on the chat benchmark.}
  \label{fig_Model_Responses_of_Chat_Benchmark}
\end{figure*}

In Figure~\ref{fig_Model_Responses_of_Foundation_Benchmark}, we illustrate a representative response from various models on the foundation benchmark. The upper portion of the figure displays the question along with the metadata for the corresponding audio. This metadata is not provided as input to the models under evaluation, the models only have access to the audio and the question posed. The lower two columns of the figure document the responses from the 9 models being tested. Similarly, an example of responses from various models on the chat benchmark can be seen in Figure~\ref{fig_Model_Responses_of_Chat_Benchmark}.

\section{Details in Human Evaluation}
We conducted a pairwise crowd worker evaluation to assess the alignment between the judgments derived from GPT-4 and those of human evaluators for both the foundation and chat benchmarks. Each pair of evaluations was scrutinized by three native English-speaking judges. During the evaluation process, we required that the entire test be conducted in a quiet environment, with human evaluators wearing headphones to listen to the audio and to isolate noise. After obtaining the test results, we conducted sample feedback; if we identified any instances of erroneous annotations, we would report back to the outsourcing platform for them to carry out a re-evaluation.

\begin{figure*}[h]
\includegraphics[width=\textwidth]{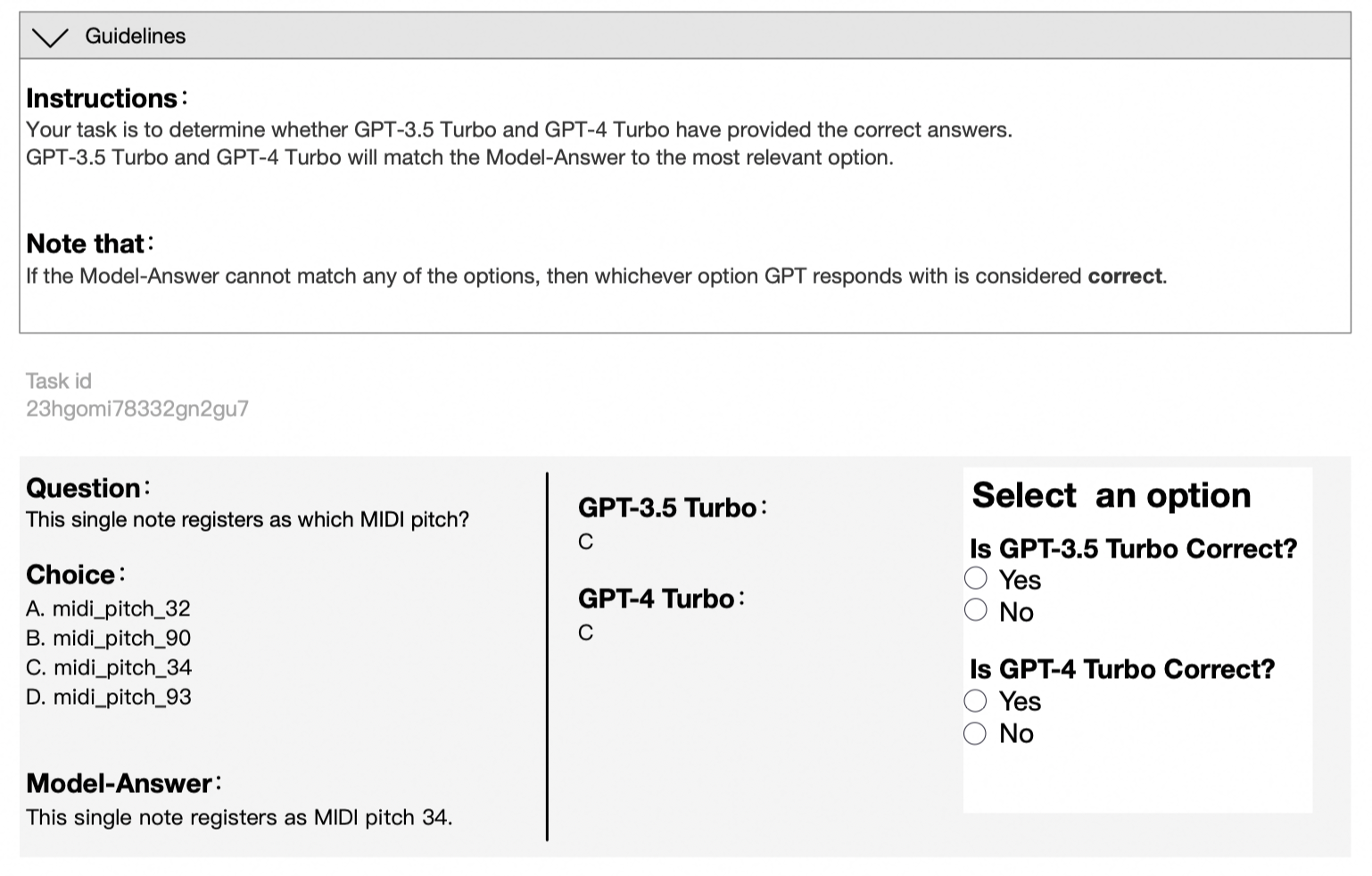}
\caption{Screenshot of human evaluation for the foundation benchmark.}
\label{figure_Screenshot_foundation}
\end{figure*}

\begin{figure*}[h]
\includegraphics[width=\textwidth]{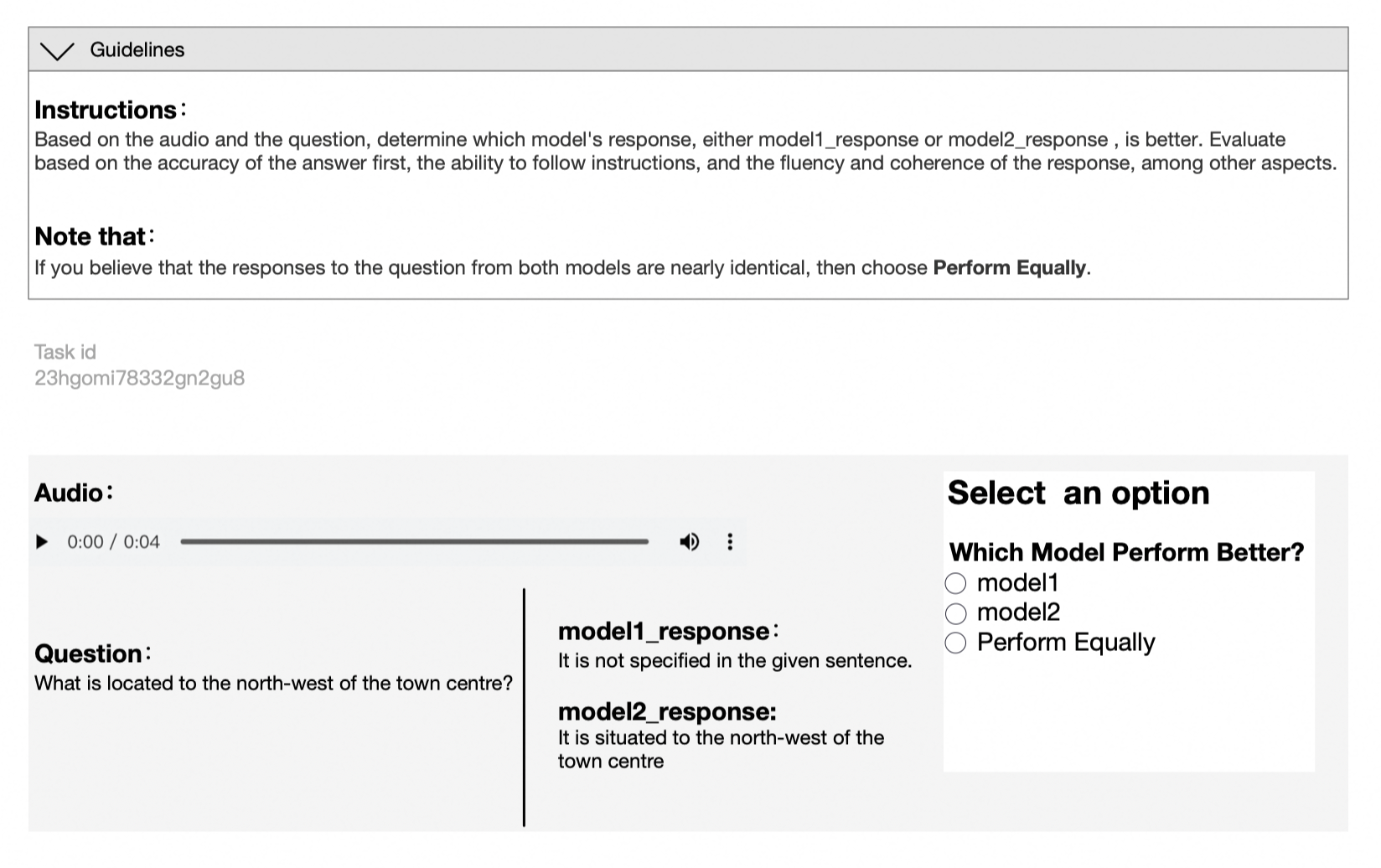}
\caption{Screenshot of human evaluation for the chat benchmark.}
\label{figure_Screenshot_chat}
\end{figure*}

\begin{itemize}
    \item For the foundation benchmark, we randomly selected 400 questions from the pool of model responses. These were accompanied by both GPT-3.5 and GPT-4 alignment results. Evaluators were instructed to ascertain whether the responses provided by GPT-3.5 Turbo and GPT-4 Turbo was accurate. The screenshots of instructions for the foundation benchmark is shown in Figure~\ref{figure_Screenshot_foundation}.
    \item  For the chat benchmark, we randomly chose 200 dialogues from the responses generated by Qwen-Audio-Chat, SALMONN, BLSP, and GPT-4, respectively. Evaluators were tasked with determining which model exhibited superior or equivalent performance. The screenshots of instructions for the chat benchmark is shown in Figure~\ref{figure_Screenshot_chat}.
    \item For the chat benchmark, we further analyzed correlation with human judgment based on task and audio type. After conducting a statistical analysis of the randomly selected QA pairs, we found that Speech accounts for 42\%, Sound for 22\%, Music for 16\%, and Mixed Audio for 20\%. To further confirm the association between human judgment and audio type, we categorized the results from Figure~\ref{fig_three_figures}(b) by audio type. As shown in Table~\ref{Table:association_of_chat}, the statistical results presented in the table indicate that QAs involving Music and Mixed Audio categories tend to have slightly higher alignment most of the time, whereas QAs involving Sound and Speech categories tend to have slightly lower alignment most of the time. We speculate that the reasons for the discrepancies might be: there are many situational questions in the Sound category QAs (such as 'What would you do if you heard this sound'), and many reasoning questions in the Speech category QAs. These more complex questions pose relatively greater challenges for GPT's evaluation. 
\end{itemize}

\begin{table*}[]
\resizebox{\textwidth}{!}{%
\begin{tabular}{clll}
\toprule
\textbf{Types} & \textbf{Task} & \textbf{Question Example} & \textbf{Choice Example} \\ \hline
\multirow{9}{*}{\textbf{Speech}} & Speech Grounding & Choose when `hate' is spoken. & \begin{tabular}[c]{@{}l@{}}A.{[}7.67, 8.05{]}  B.{[}1.03, 1.53{]}\\ C.{[}3.07, 3.27{]}  D.{[}7.02, 7.21{]}\end{tabular} \\ \cline{2-4} 
 & Spoken language identification & Recognize the language of the speech. & A.en  B.ja  C.de  D.fr \\ \cline{2-4} 
 & \begin{tabular}[c]{@{}l@{}}Speaker gender recognition\\ (biologically)\end{tabular} & Detect the gender of the speaker in this audio file. & A.male B.female \\ \cline{2-4} 
 & Emotion recognition & What emotion is at the forefront of the speaker's words? & \begin{tabular}[c]{@{}l@{}}A.angry  B.happy\\ C.sad      D.neutral\end{tabular} \\ \cline{2-4} 
 & Speaker age prediction & Which age range do you believe best matches the speaker's voice? & \begin{tabular}[c]{@{}l@{}}A.teens to twenties \\ B.thirties to forties\\ C.fifties to sixties\\ D.seventies to eighties\end{tabular} \\ \cline{2-4} 
 & Speech entity recognition & Tell me the first `transport\_type'-connected word in this audio. & \begin{tabular}[c]{@{}l@{}}A.go       B.how\\ C.metro  D.train\end{tabular} \\ \cline{2-4} 
 & Intent classification & What's your opinion on the speaker's goal in this sound clip? & \begin{tabular}[c]{@{}l@{}}A.audio\_volume\_up\\ B.news\_query\\ C.lists\_createoradd\\ D.play\_podcasts\end{tabular} \\ \cline{2-4} 
 & Speaker number verification & The speech features how many speakers? & A.2  B.4  C.3  D.1 \\ \cline{2-4} 
 & Synthesized voice detection & Based on your assessment, is this speech Real or Fake? & A.fake  B.real \\ \hline
\multirow{4}{*}{\textbf{Sound}} & Audio grounding & \begin{tabular}[c]{@{}l@{}}What are the exact times when `a woman briefly talks' is \\ present in the clip?\end{tabular} & \begin{tabular}[c]{@{}l@{}}A.{[}0.44, 2.38{]}\\ B. {[}3.85, 4.11{]}\\ C. {[}9.01, 10.02{]}\\ D. {[}4.15, 7.83{]}\end{tabular} \\ \cline{2-4} 
 & Vocal sound classification & What's the provenance of the sound in this clip? & \begin{tabular}[c]{@{}l@{}}A.Sigh     B.Throat clearing\\ C.Cough  D.Sneeze\end{tabular} \\ \cline{2-4} 
 & Acoustic scene classification & What venue are the sounds indicative of? & \begin{tabular}[c]{@{}l@{}}A.kitchen  B.elevator\\ C.street  D.crowded indoor\end{tabular} \\ \cline{2-4} 
 & Sound question answering & What animal makes a sound in the video? & \begin{tabular}[c]{@{}l@{}}A.cattle  B.horse\\ C.cat       D.bird\end{tabular} \\ \hline
\multirow{6}{*}{\textbf{Music}} & Music instruments classification & Discern the principal instrument in this tune. & \begin{tabular}[c]{@{}l@{}}A.bass   B.string\\ C.brass  D.mallet\end{tabular} \\ \cline{2-4} 
 & Music genre classification & What's the genre identity of this music? & \begin{tabular}[c]{@{}l@{}}A.Jazz       B.Rock\\ C.Country D.Experimental\end{tabular} \\ \cline{2-4} 
 & Music note analysis-pitch & What is the MIDI pitch level of the note played? & \begin{tabular}[c]{@{}l@{}}A.midi\_pitch\_19\\ B.midi\_pitch\_29\\ C.midi\_pitch\_37\\ D.midi\_pitch\_71\end{tabular} \\ \cline{2-4} 
 & Music note analysis-velocity & What numerical value is the MIDI velocity for this note? & \begin{tabular}[c]{@{}l@{}}A.midi\_velocity\_127\\ B.midi\_velocity\_50\\ C.midi\_velocity\_100\\ D.midi\_velocity\_25\end{tabular} \\ \cline{2-4} 
 & Music question answering & Is the guzheng louder than the piano? & A.yes  B.no  C.four  D.one \\ \cline{2-4} 
 & Music emotion detection & What kind of sentiment does this music invoke? & \begin{tabular}[c]{@{}l@{}}A.meditative  B.positive\\ C.trailer          D.advertising\end{tabular} \\ 
\bottomrule
\end{tabular}%
}
\caption{Examples of questions and choices on the foundation benchmark.}
\label{Table_Examples_of_Foundation}
\end{table*}

\begin{table*}[]
\resizebox{\textwidth}{!}{%
\begin{tabular}{c|cccccccc}
\toprule
\textbf{Categories} & \textbf{Qwen-Audio} & \textbf{Qwen-Audio Turbo} & \textbf{SALMONN} & \textbf{BLSP} & \textbf{NExT-GPT} & \textbf{SpeechGPT} & \textbf{PandaGPT} & \textbf{Whisper+GPT-4} \\ \hline
\textbf{Speech grounding} & \textbf{56.1\%} & 45.4\% & 25.3\% & 25.0\% & 25.4\% & 28.8\% & 23.0\% & 35.0\% \\ \hline
\textbf{\begin{tabular}[c]{@{}c@{}}Spoken language\\ identification\end{tabular}} & 92.8\% & 95.9\% & 28.1\% & 30.8\% & 23.7\% & 39.6\% & 34.6\% & \textbf{96.8\%} \\ \hline
\textbf{\begin{tabular}[c]{@{}c@{}}Speaker gender\\ recognition\end{tabular}} & 67.2\% & \textbf{82.5\%} & 35.5\% & 33.2\% & 57.0\% & 29.2\% & 66.5\% & 21.9\% \\ \hline
\textbf{Emotion recognition} & 43.2\% & \textbf{60.0}\% & 29.9\% & 27.4\% & 25.7\% & 37.6\% & 26.0\% & 59.5\% \\ \hline
\textbf{\begin{tabular}[c]{@{}c@{}}Speaker age\\ prediction\end{tabular}} & 36.0\% & 58.8\% & 48.7\% & 51.2\% & \textbf{62.4}\% & 20.4\% & 42.5\% & 41.1\% \\ \hline
\textbf{\begin{tabular}[c]{@{}c@{}}Speech entity\\ recognition\end{tabular}} & \textbf{71.2\%} & 48.1\% & 51.7\% & 37.2\% & 26.1\% & 35.9\% & 34.0\% & 69.8\% \\ \hline
\textbf{Intent classification} & 77.8\% & 56.4\% & 36.7\% & 46.6\% & 25.6\% & 45.8\% & 28.5\% & \textbf{87.7\%} \\ \hline
\textbf{\begin{tabular}[c]{@{}c@{}}Speaker number\\  verification\end{tabular}} & 35.3\% & \textbf{54.3\%} & 34.3\% & 28.1\% & 25.4\% & 32.6\% & 43.2\% & 30.0\% \\ \hline
\textbf{\begin{tabular}[c]{@{}c@{}}Synthesized voice\\ detection\end{tabular}} & 48.3\% & \textbf{69.3\%} & 50.0\% & 50.0\% & 30.8\% & 39.2\% & 53.1\% & 40.5\% \\ \hline
\textbf{Audio grounding} & 23.9\% & 41.6\% & 24.0\% & 34.6\% & \textbf{62.2\%} & 26.1\% & 38.3\% & / \\ \hline
\textbf{\begin{tabular}[c]{@{}c@{}}Vocal sound\\ classification\end{tabular}} & \textbf{84.9\%} & 78.1\% & 45.3\% & 29.8\% & 23.5\% & 26.2\% & 31.6\% & / \\ \hline
\textbf{\begin{tabular}[c]{@{}c@{}}Acoustic scene\\ classification\end{tabular}} & \textbf{67.5\%} & 61.3\% & 34.1\% & 25.2\% & 24.1\% & 23.7\% & 55.7\% & / \\ \hline
\textbf{\begin{tabular}[c]{@{}c@{}}Sound question\\ answering\end{tabular}} & \textbf{64.6\%} & 62.8\% & 28.4\% & 36.1\% & 18.8\% & 33.9\% & 48.7\% & / \\ \hline
\textbf{\begin{tabular}[c]{@{}c@{}}Music instruments\\ classification\end{tabular}} & 59.1\% & \textbf{59.6\%} & 41.3\% & 22.8\% & 24.3\% & 29.1\% & 47.7\% & / \\ \hline
\textbf{\begin{tabular}[c]{@{}c@{}}Music genre \\ classification\end{tabular}} & 71.2\% & \textbf{77.1\%} & 45.3\% & 26.1\% & 28.1\% & 29.3\% & 39.8\% & / \\ \hline
\textbf{\begin{tabular}[c]{@{}c@{}}Music note \\ analysis-pitch\end{tabular}} & 28.6\% & \textbf{30.1\%} & 26.4\% & 23.5\% & 25.1\% & 24.1\% & 26.4\% & / \\ \hline
\textbf{\begin{tabular}[c]{@{}c@{}}Music note\\ analysis-velocity\end{tabular}} & 25.4\% & 25.1\% & 22.8\% & 24.9\% & 23.1\% & 25.2\% & \textbf{27.2\%} & / \\ \hline
\textbf{\begin{tabular}[c]{@{}c@{}}Music question \\ answering\end{tabular}} & 48.2\% & \textbf{62.5\%} & 54.6\% & 31.0\% & 47.1\% & 31.3\% & 50.7\% & / \\ \hline
\textbf{Music emotion detection} & 36.1\% & \textbf{39.0\%} & 32.2\% & 28.3\% & 25.4\% & 29.7\% & 36.7\% & / \\ 
\bottomrule
\end{tabular}%
}
\caption{The accuracy of each model across all tasks in the foundation benchmark.}
\label{Table_Foundation_Results}
\end{table*}

\begin{table*}[]
\begin{tabular}{l|llllll}
\toprule
\textbf{Type} & \textbf{\begin{tabular}[c]{@{}l@{}}GPT-4 vs\\ BLSP\end{tabular}} & \textbf{\begin{tabular}[c]{@{}l@{}}GPT-4 vs \\ Qw.Chat\end{tabular}} & \textbf{\begin{tabular}[c]{@{}l@{}}GPT-4 vs\\ SALMONN\end{tabular}} & \textbf{\begin{tabular}[c]{@{}l@{}}SALMONN\\ vs BLSP\end{tabular}} & \textbf{\begin{tabular}[c]{@{}l@{}}SALMONN\\ vs Qw.Chat\end{tabular}} & \textbf{\begin{tabular}[c]{@{}l@{}}Qw.Chat \\ vs BLSP\end{tabular}} \\ \hline
\textbf{Speech} & 77\% & 76\% & 89\% & 73\% & 75\% & 69\% \\
\textbf{Sound} & 73\% & 66\% & 96\% & 66\% & 75\% & 73\% \\
\textbf{Music} & 75\% & 88\% & 88\% & 81\% & 84\% & 75\% \\
\textbf{Mixed Audio} & 83\% & 88\% & 93\% & 75\% & 78\% & 70\% \\ 
\bottomrule
\end{tabular}%
\caption{Association between human judgment and audio type.}
\label{Table:association_of_chat}
\end{table*}

\end{document}